# Risk Factors Associated with Mortality in Game of Thrones: A Longitudinal Cohort Study

Suveen Angraal MBBS,* Ambika Bhatnagar MBBS,* Suraj Verma MBBS,
Sukhman Shergill MBBS, Aakriti Gupta MD and Rohan Khera MD

**ABSTRACT**

**Objective:** To assess mortality, and identify the risk factors associated with mortality in Game of Thrones (GoT).
**Design and Setting:** A longitudinal cohort study in the fictional kingdom of Westeros and Essos.
**Participants:** All the characters appearing in the GoT since airing of its first episode with screen time of ≥5 minutes.
**Main Outcome Measures:** All-cause mortality. Multivariate Cox proportional hazard model was used to assess the risk factors associated with mortality, represented by hazard ratios, with episode as the unit of time.
**Results:** Of the 132 characters, followed up for a median time of 32 episodes, a total 89 (67.4%) characters died; with external invasive injury as the most common cause of death, attributing to 42.4% of the total deaths. Age (in decades) was a significant risk factor for death [HR, 1.24 (95% CI, 1.08-1.43), P=0.0001]. Although statistically non-significant, allegiance to house Targaryen [HR, 1.10 (95% CI, 0.32-3.77)] was associated with a higher risk for mortality per episode than house Stark. Characters residing in South were less likely to die than characters residing in North [HR, 0.58 (95% CI, 0.29-1.16), P=0.12]. Advisors showed a lower risk of mortality than the members of houses, with some statistical significance [HR, 0.39 (95% CI, 0.14-1.08), P=0.07].
**Conclusions:** There is a high mortality rate among the characters in GoT. Residing in the North and being a member of a house is very dangerous in GoT. Allegiance to house Stark trended to be safer than house Targaryen.

## BACKGROUND

Game of Thrones (GoT) has emerged as the frontrunner of contemporary entertainment, and is one of the most viewed television series in recent years, with the finale of seventh season garnering approximately 12.1 million viewers globally.[1] The series is known to feature gut-churning violence. While television series are usually forgiven for high mortality due to their artistic license,[2] GoT has continued to surprise its viewers by killing off protagonist characters. Ardent viewers of the show, including the authors of the present study, have frequently been challenged by the inability to predict the impending demise of the remaining characters. After many emotional turbulences on witnessing several poignant and grotesque deaths of their beloved characters, the authors sought to pursue a scientific investigation into understanding the pattern of mortality in GoT, and the predictors of death on the show.

## METHODS

Data was collected using the HBO Go online repository starting from the first episode which aired on April 17, 2011. The study cohort included all the characters appearing in the GoT since airing of its first episode, who garnered a total screen time ≥5 minutes. Baseline characteristics, including age, sex, allegiance, occupation and geographical residence of the characters were assessed at their first appearance on the show.

For the purpose of this study, we divided the study population into 7 major houses, based on allegiance. These included houses Stark, Targaryen, Lannister, Baratheon, Greyjoy, Martell and Tyrell. Allegiance to any house was defined through family name of the character. If the character does not belong to any of the major house families, the allegiance was assessed on the basis of the pledged association towards one of the major houses. Occupation was assessed on the basis of primary responsibility. '*Member of a house*' was assigned to a character if he/she is born or adopted

From the Center for Outcomes Research and Evaluation, Yale-New Haven Health, New Haven, Connecticut (S.A), Department of Pediatrics Emergency Medicine, Yale School of Medicine, New Haven, Connecticut (A.B), Department of Psychiatry, All India Institute of Medical Sciences, New Delhi, India (S.V). All India Institute of Medical Sciences, New Delhi, India (S.S), Division of Cardiology, Columbia University Medical Center, New York, New York (A.G), Division of Cardiology, UT Southwestern Medical Center, Dallas, TX (R.K). *Contributed equally.

**Corresponding Author**
Dr. Suveen Angraal; 55 Church Street, Suite 401, New Haven, CT 06510
+1-203-737-2090; suveen.angraal@yale.edu

in any of the major houses. A character was classified as an *advisor* or *knight/soldier* if he/she is not a member of a house, but serves the primary role of an advisor to house members or the role of a soldier respectively. GoT takes place in fictional land of the 'Known World' with epicenter in the kingdom of Westeros.[3] For geographic distribution, a character was considered a resident of the North if he/she resided north of the River Road, a major highway running west to east in Westeros which bisects the kingdom.[4] Characters living at any location south of the River Road were considered residents of the South. Essos was the region east of the Narrow Sea.

All-cause mortality was primary outcome of interest. The temporal progression of the show was measured in the numerical order of the episodes, and causes of death were noted for all the characters on the show. We excluded the supernatural characters that did not age. If a death of a character occurred at the hands of a 'White Walker', the character was excluded from the cohort; since that can result in zombification of the person, who technically dead, may continue to be a character on the show. The characters were then followed up until the airing of last (67th) episode of seventh season of the show. Data were supplemented through other online resources, and missing data were completed for all the study participants. Data was sourced by two authors (SV and SS), and confirmed by a third (SA).

Survival was analyzed using Kaplan-Meier survival curves. Cox proportional hazard model was used to assess the risk factors associated with mortality, represented by hazard ratios. We chose one episode as the unit of time, since the elapsed time is uneven in television shows. However, we also performed sensitivity analyses using logistic regression model to estimate the odds of death by the end of the follow up period. The analysis was conducted using R programming language version 3.3.1.

**RESULTS**
Our study included 132 characters, followed up for a median time of 32 episodes. Baseline characteristics of the study cohort are included in the **Table 1**. Mean age of the study population was 35.1 years. Majority of the characters belonged to the house Stark (19.7%), followed closely by houses Lannister (14.4%) and Targaryen (10.6%). Out of the 132 characters, 89 (67.4%) characters died during follow-up. Survival varied among houses, with house Lannister showing a better survival than houses Stark and Targaryen for most of the study period (**Figure 1**). The mode of death for most of the characters was external injury, and only 1 natural death. A total of 59 (44.7%) characters died due to some form of invasive injury, while 12 (9.1%) died from severe burns, and 4 (3.0%) due to ingestion of poisonous materials.

Results from the multivariable Cox proportional hazards model are shown in **Table 2**. Age (in decades) was a significant risk factor for death [HR, 1.24 (95% CI, 1.08-1.43), P=0.009], while sex was not [HR of male gender, 1.02, (95% CI, 0.56-1.86)]. Although not statistically significant, allegiance to house Targaryen was associated with a higher risk for mortality per episode than house Stark; HR, 1.10 (95% CI, 0.32-3.77). Further, allegiance to houses Baratheon, Martell, or Tyrell was also associated with higher risk of mortality as compared to house Stark. Characters residing in South were less likely to die than characters residing in North [HR, 0.58 (95% CI, 0.29-1.16), P=0.12]. Advisors showed a lower risk of mortality than the members of houses, with some statistical significance [HR, 0.39 (95% CI, 0.14-1.08), P=0.07].

Multivariable logistic regression model complemented the results we obtained from the Cox proportional hazards model (**Table 2**).

**DISCUSSION**
This is the first scientific investigation to assess predictors of mortality for a world-famous TV series, GoT. In this longitudinal study, we noted that Westeros is a very dangerous place to live in, with more than two-thirds of the characters dead by the end of study period. The gut-wrenching violence featured in the series is further evident by the fact that majority of the characters died due to external invasive injury. Majority of the houses showed a trend towards higher risk of mortality than Starks. Furthermore, in reference to member of a house, the advisors showed lower risk of dying per episode. Finally, residing in the South was safer than residing in the North.

Researchers have been intrigued by GoT's premise, and how morbidities on the show can be congruous to real word.[5] Previous studies have shown that the TV dramas have high mortality rates.[2] Our study extends the previous work, and assesses the risk factors associated with mortality in one of the most famous television shows of all times.

In comparison to house Stark, the house with the majority of protagonists of the show, the allegiance to most of the other houses showed a higher risk for mortality. Although statistically non-significant, this bodes well for the

'*current*' protagonist- Jon Snow, and may clue towards an unfavorable outcome for Daenerys Targaryen, the other protagonist of the show. Given that the author of the show has planned an ending where one of the protagonist will die,[6] our results steer towards the probability of a Targaryen death.

Our study has limitations. First, given that it is difficult to assess time in a fantasy series, the time in our analysis was defined in episodes. Nonetheless, results can still be interpreted to fulfill the aim of the study. Furthermore, we performed sensitivity analysis using logistic regression, and results were in agreement with results from Cox model. Secondly, we were not able to account for change in allegiance of some characters, as the series progressed. However, most of characters did not show any significant change in characteristics over study period. Finally, this is an observational study with a small sample size on subjects which are fictional. However, the current study design allows us to draw conclusions on common characteristics of the subjects, and how these characteristics relate to mortality.

In conclusion, the characters in GoT suffer from extremely high mortality rate. Residing in the North and being a member of a house is very dangerous in GoT. Divergent to popular belief, allegiance to house Stark trended to be safer than most of the houses.

We sincerely apologize to the producers of the show for any potential spoilers that came out of our study. We hope that our study will help the fans of this show to be prepared for what is to come. We also hope our study will incite critical thinking among the readers, encouraging them to take a more scientific approach to daily questions.

*Valar Morghulis!*

## REFERENCES

**1.** John Koblin, The New York Times. 'Game of Thrones' Finale Sets Ratings Record. 2017;
https://www.nytimes.com/2017/08/28/arts/television/game-of-thrones-finale-sets-ratings-record.html. Accessed 15 November, 2017.

**2.** Crayford T, Hooper R, Evans S. Death rates of characters in soap operas on British television: is a government health warning required? *BMJ.* 1997;315(7123):1649-1652. doi: https://doi.org/10.1136/bmj.315.7123.1649.

**3.** Interactive Game of Thrones Map. https://quartermaester.info/. Accessed 8 January, 2018.

**4.** Game of Thrones Wiki. River Road. http://gameofthrones.wikia.com/wiki/River_Road. Accessed 8 January, 2018.

**5.** Lipoff JB. Greyscale—a mystery dermatologic disease on hbo's game of thrones. *JAMA Dermatology.* 2016;152(8):904-904. doi: 10.1001/jamadermatol.2015.5793.

**6.** Ashley Ross, TIME. George R. R. Martin Tells Fans To Expect 'Bittersweet' Ending to Game of Thrones. 2015; http://time.com/4101276/game-of-thrones-ending-george-r-r-martin/.
Accessed 12 December, 2017.

**Disclosures:** Dr. Gupta is supported by grant HL007854 from the National Institutes of Health. Dr. Khera is supported by the National Heart, Lung, and Blood Institute (5T32HL125247-02) and the National Center for Advancing Translational Sciences (UL1TR001105) of the National Institutes of Health. Other authors have no disclosures to declare. The sponsors were not involved in the design and conduct of this study.

**Table 1. Baseline Characteristics of the Study Population.**

| Characteristics | | |
|---|---|---|
| **Age, years [mean(SD)]** | 35.1 (18.3) | |
| | **Study Population (n=132)** | **Deaths [n(%)]** |
| **Male (%)** | 100 (75.8) | 68 (66.0) |
| **Female (%)** | 32 (24.2) | 21 (65.6) |
| **Allegiance (%)** | | |
| Stark | 26 (19.7) | 13 (50.0) |
| Targaryen | 14 (10.6) | 9 (64.3) |
| Lannister | 19 (14.4) | 11 (57.9) |
| Baratheon | 13 (9.8) | 10 (76.9) |
| Greyjoy | 5 (3.8) | 2 (40.0) |
| Martell | 6 (4.5) | 5 (83.3) |
| Tyrell | 4 (3.0) | 4 (100.0) |
| Other Allegiance/Commoners | 45 (34.1) | 35 (77.8) |
| **Occupation (%)** | | |
| Member of a House or Royalty | 42 (31.8) | 29 (69.0) |
| Knight or Soldier | 46 (34.8) | 33 (71.7) |
| Advisor | 13 (9.8) | 7 (53.8) |
| Other Occupation | 31 (23.5) | 20 (64.5) |
| **Geographic Location (%)** | | |
| North | 59 (44.7) | 42 (71.2) |
| South | 49 (37.1) | 31 (63.3) |
| Essos | 24 (18.2) | 16 (66.7) |
| **Total** | | 89 (67.4) |

**Table 2. Results from Cox Proportional Hazard Model and Logistic Regression Model**

|  | Multivariate Cox Model | | | Multivariate Logistic Regression Model | |
| --- | --- | --- | --- | --- | --- |
|  | Coefficient | Hazard Ratio (95% CI) | P value | Odds Ratio (95% CI) | P value |
| **Age(y)/10** | 0.22 | 1.24 (1.08-1.43) | 0.0001 | 1.06 (1.01-1.12) | 0.01 |
| **Male** | 0.02 | 1.02 (0.56-1.86) | 0.96 | 0.98 (0.80-1.20) | 0.83 |
| **Allegiance** (Reference: House stark) | | | | | |
| Baratheon | 0.6 | 1.74 (0.64-4.73) | 0.27 | 1.67 (1.16-2.40) | 0.01 |
| Greyjoy | -0.62 | 0.54 (0.11-2.54) | 0.43 | 0.82 (0.53-1.28) | 0.38 |
| Lannister | -0.10 | 0.90 (0.34-2.44) | 0.84 | 1.21 (0.88-1.65) | 0.24 |
| Martell | 1.70 | 5.46 (1.44-20.71) | 0.01 | 1.74 (1.07-2.82) | 0.03 |
| Targaryen | 0.09 | 1.10 (0.32-3.77) | 0.88 | 1.33 (0.89-2.00) | 0.17 |
| Tyrell | 0.59 | 1.80 (0.43-7.63) | 0.42 | 1.92 (1.11-3.32) | 0.02 |
| Other Allegiance | 0.52 | 1.69 (0.83-3.44) | 0.15 | 1.32 (1.03-1.69) | 0.03 |
| **Occupation** (Reference: Member of a House) | | | | | |
| Advisor | -0.93 | 0.39 (0.14-1.08) | 0.07 | 0.80 (0.57-1.11) | 0.18 |
| Knight or Soldier | -0.08 | 1.09 (0.47-1.81) | 0.81 | 0.99 (0.78-1.26) | 0.95 |
| Other Occupation | 0.17 | 1.19 (0.53-2.68) | 0.68 | 0.98 (0.75-1.29) | 0.90 |
| **Geographic Location** (Reference: 'North') | | | | | |
| South | -0.55 | 0.58 (0.29-1.16) | 0.12 | 0.75 (0.59-0.96) | 0.02 |
| Essos | 0.23 | 1.26 (0.51-3.14) | 0.61 | 0.85 (0.63-1.14) | 0.28 |

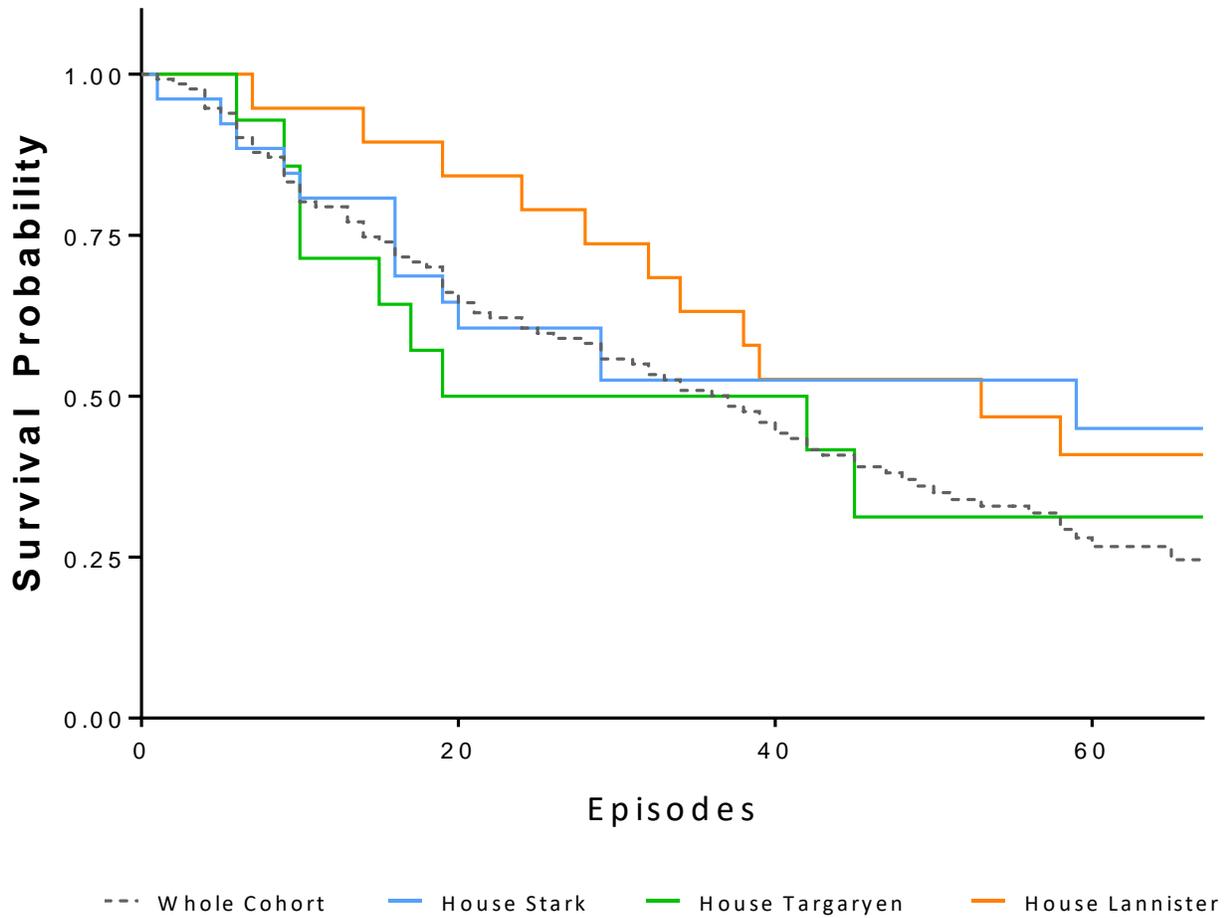

Figure 1. Survival in Houses Stark, Targaryen and Lannister